\documentclass[aps,twocolumn,10pt,nofootinbib,groupedaddress]{revtex4-1}
\usepackage{amssymb,amsmath,amstext,amsfonts}
\usepackage{bbold}
\usepackage{bm}
\usepackage{graphicx}
\usepackage{xcolor}
\usepackage[normalem]{ulem}
\usepackage{soul}

\newcommand{\beq}{\begin{equation}}
\newcommand{\eeq}{\end{equation}}
\newcommand{\bal}{\begin{aligned}}
\newcommand{\eal}{\end{aligned}}
\newcommand{\Lag}{{\mathcal{L}}}
\newcommand{\Mp}{M_{\rm Pl}}

\newcommand{\etaperp}{\eta_\perp}
\newcommand{\F}{Q_s}
\newcommand{\etamax}{\eta_\perp^{\textrm{max}}}
\newcommand{\Nf}{N_{\textrm{f}}}
\newcommand{\sig}{\Delta}
\newcommand{\Nbar}{\bar{N}}
\newcommand{\kp}{k_{\textrm{p}}}
\newcommand{\kf}{k_{\textrm{f}}}
\newcommand{\ke}{k_{\textrm{e}}}

\begin{document}

\title{Turning in the landscape: \\ a new mechanism for generating Primordial Black Holes} 

\author{Jacopo Fumagalli, S\'ebastien Renaux-Petel, John W.~Ronayne and Lukas T.~Witkowski}
\affiliation{Institut d'Astrophysique de Paris, GReCO, UMR 7095 du CNRS et de Sorbonne Universit\'e,
98bis boulevard Arago, Paris 75014, France}

\date{\today}

\begin{abstract}
We propose a new model-independent mechanism for producing primordial black holes from a period of multi-field inflation. 
This requires an enhancement of primordial fluctuations at short scales compared to their value at CMB scales. We show that such an amplification naturally occurs when the inflationary trajectory exhibits a strong turn, that is a limited period during which the trajectory strongly deviates from a geodesic in field space, and is sufficient for
subsequently producing primordial black holes with
the abundance to be all or a fraction of dark matter.
Our mechanism is generic to models of inflation realized in a multi-dimensional field space with an overlying potential and geometry, also referred to as the inflationary landscape, as arises in embeddings of inflation in high-energy theories. We study analytically and numerically how the duration and the strength of the turn impact the primordial fluctuation power spectrum and the abundance of primordial black holes. Our mechanism has the potential of exhibiting unique features accessible to observation through the primordial black hole spectrum and the stochastic background of gravitational waves, offering a precious glimpse at the dynamics of inflation.
 
\end{abstract}

\maketitle


{\bf Introduction.}--- The nature of Dark Matter (DM) is a great riddle of contemporary fundamental physics. A possibility that has been entertained for some time is that all or a fraction of DM can consist of primordial black holes (PBHs) \cite{Chapline:1975ojl}, i.e.~black holes resulting from the collapse of local overdensities in the early universe.
The idea that quantum fluctuations during cosmological inflation are responsible for seeding the overdensities responsible for PBH creation has been contemplated since the early work of \cite{Ivanov:1994pa, 9605094, 9611106}. A particular challenge is that the amplitude of fluctuations needs to be larger by a factor of $\sim 10^7$ compared to the amplitude observed at CMB scales $\mathcal{P}_\zeta \sim 10^{-9}$ to lead to a significant production of PBHs. In models of single-field inflation this can be achieved by invoking suitable features in the single field Lagrangian \cite{0711.5006, 1606.07631, 1702.03901, 1706.04226, 1706.06784, 1709.05565, 1712.09750,Cai:2018tuh,Ballesteros:2018wlw,Cai:2019bmk} or by coupling the inflation to gauge fields \cite{1212.1693, 1312.7435, 1704.03464}. 
Considering two stages of inflation driven by different fields naturally
offers even more flexibility for model-building \cite{Kawasaki:2016pql,Inomata:2017okj}. Enhanced fluctuations can also arise from inflating on a ridge, like in the waterfall phase in hybrid inflation \cite{9605094, 1107.1681, 1112.5601, 1501.07565, 1512.03515}.

In this letter, we present a new PBH production mechanism based on a novel way of enhancing fluctuations exploiting multi-field dynamics and with unique observational signatures. Motivated by embeddings of inflation in high-energy physics, a promising arena for model-building involves
inflationary trajectories characterized by substantial 
departures from geodesics in a multi-dimensional
field space, i.e.~strong turns that are ubiquitous when the former is negatively curved~\cite{Cremonini:2010ua,Renaux-Petel:2015mga,1607.08609,Brown:2017osf,Mizuno:2017idt,1803.09841,Garcia-Saenz:2018ifx,Achucarro:2018vey,Bjorkmo:2019aev,1902.10529,Christodoulidis:2019jsx,Aragam:2019khr,Bravo:2019xdo,Garcia-Saenz:2019njm,Chakraborty:2019dfh,Renaux-Petel:2021yxh}. In this context we show that a limited phase of strongly non-geodesic motion can enhance the curvature fluctuation power spectrum sufficiently to allow for subsequent PBH formation. The mechanism is not restricted to a particular realization of inflation, and we explain it model-independently in terms of the masses and couplings governing the physics of fluctuations. Central to our mechanism is the time dependence of the
``bending'' parameter $\etaperp$ measuring the strength of geodesic deviation, and we contrast the situations of broad and sharp turns, and highlight the differences between our mechanism and the well known one of hybrid inflation. We explain how the time-dependence of $\etaperp$ enables one to characterize the peak of the power spectrum, finding in particular that its growth can overcome the bound deduced in the single-field case \cite{Byrnes:2018txb, Carrilho:2019oqg,Ozsoy:2019lyy}. We further analyze how the model parameters affect the mass spectrum of PBHs, assuming Gaussian statistics for the fluctuations. 
The inherent sensitivity of the PBH abundance on departures from Gaussian statistics (see \cite{1206.4188,1307.4995,1801.09415,1811.07857,DeLuca:2019qsy,1906.02827,1906.06790,1908.11357,1912.05399}) leaves us with the exciting prospect that the distinctive pattern of non-Gaussianities (NG) tied to the mechanism might imprint characteristic features in the PBH spectrum. Another telltale signature of this mechanism, for turns executed in less than one $e$-fold, are oscillations in $\mathcal{P}_\zeta$, which lead to a modulated spectrum of induced gravitational waves (GWs) \cite{Fumagalli:2020nvq,Braglia:2020taf,Fumagalli:2021mpc}, which is potentially observable in near-future GW observatories \cite{Caprini:2019pxz,Pieroni:2020rob,Flauger:2020qyi,Fumagalli:2021dtd}. A detection would offer invaluable insights into the dynamics behind inflation.


{\bf PBH generation mechanism.}--- Our starting point is the generic two-derivative action for scalar fields $\phi^I$ minimally coupled to gravity:
\beq \label{eq:action1}
S=\int d^4x\sqrt{-g}\bigg[\frac{\Mp^2}{2}\,R-\frac{1}{2}\,G_{IJ}\nabla^{\mu}\phi^I\nabla_{\mu}\phi^J-V(\phi)\bigg]\,,
\eeq
where $G_{IJ}(\phi)$ defines a metric in the internal field space parametrized by the coordinates $\phi^I$. The inflationary background dynamics is characterized by a spatially flat 
Friedmann-Lema\^itre-Robertson-Walker metric with scale factor $a(t)$, Hubble parameter $H(t)=\dot{a}/a$, and homogeneous scalar fields whose equations of motion read ${\cal D}_t \dot{\phi}^I  +3H  \dot{\phi}^I +G^{IJ} V_{,J}=0$, where the time field space covariant derivative of any field space vector $A^I$ is defined as ${\cal D}_t A^I=\dot{A}^I+\Gamma^I_{JK}\dot{\phi}^JA^K$. For later convenience we define 
$\dot{\sigma} \equiv (G_{IJ}\dot{\phi}^I\dot{\phi}^J)^{1/2}$ and $\epsilon \equiv -\dot{H}/H^2$. Considering a two-field landscape for simplicity in what follows, it is particularly useful to introduce the adiabatic-entropic orthonormal basis defined by $e^I_{\sigma}\equiv \dot{\phi}^I/\dot{\sigma}$ and $e^I_s$, which is orthogonal to $e^I_{\sigma}$, and with a definite orientation for the basis $(e^I_{\sigma},e^I_s)$. The latter evolve as
\beq
{\cal D}_t e^I_{\sigma}=H\eta_{\perp}e^I_s\,,\qquad {\cal D}_t e^I_s=-H\eta_{\perp}e^I_{\sigma}\,,
\eeq
where the dimensionless  ``bending'' parameter $\etaperp=-e_s^I V_{,I}/(H \dot \sigma)$ measures the deviation of the background trajectory from a field space geodesic \cite{GrootNibbelink:2000vx,GrootNibbelink:2001qt}. 

To describe the physics of scalar linear fluctuations about the above background, 
we work in the comoving gauge, with $\delta \phi^I=\F e_s^I$ and the spatial part of the metric reading $g_{ij}=a^2 e^{2 \zeta} \delta_{ij}$.
 The so-called comoving curvature perturbation $\zeta$ is ultimately the quantity of direct observational interest, while $\F$ embodies the genuine multi-field effects. Their quadratic action can be cast in the simple form  \cite{Sasaki:1995aw,GrootNibbelink:2001qt,Langlois:2008mn} (writing $S=\int {\rm d}t \,{\rm d}^3 x {\cal L}$)
\begin{eqnarray}
\Lag^{(2)}&=&a^3\bigg[\Mp^2\epsilon\left(\dot{\zeta}^2-\frac{(\partial \zeta)^2}{a^2}\, \right)+2\dot{\sigma}\eta_{\perp}\dot{\zeta}\F \nonumber \\
&+&\frac{1}{2}\left(\dot{\F}^2-\frac{(\partial \F)^2}{a^2}-m_s^2\F^2\right)\bigg]\,,
\label{L2}
\end{eqnarray}
where the entropic mass reads
\beq
\label{ms2}
m_s^2=V_{;ss}-H^2\eta_{\perp}^2+\epsilon H^2  \Mp^2 R_{\rm fs},
\eeq
with $V_{;ss}=e_s^I e_s^J V_{;IJ}$ the projection of the covariant Hessian of the potential along the entropic direction, and $R_{{\rm fs}}$ the field space scalar curvature. This shows that the physics of linear fluctuations of any two-field model can be described by only three functions of the number of \textit{e}-folds $N=\ln(a)$: the Hubble scale $H(N)$, like in single-field models, the entropic mass $m_s^2(N)$, and the bending $\etaperp(N)$. 

\begin{figure}[t]
	\includegraphics*[width=7.5cm]{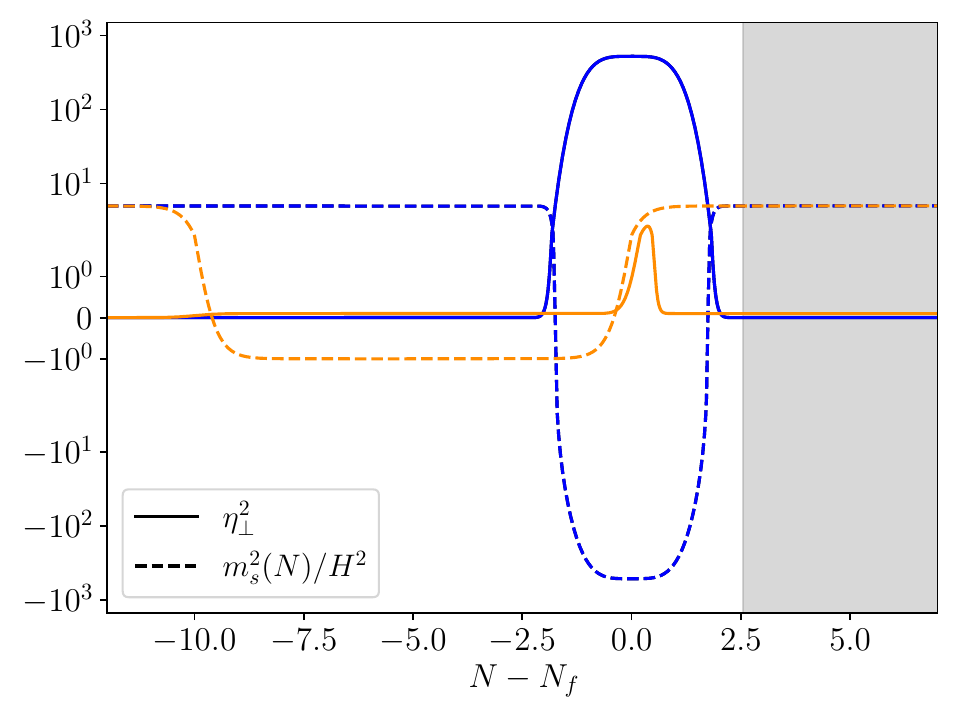}
	\caption{ 
		\textit{Schematic time dependences of the bending parameter $\etaperp^2$ (solid lines) and the entropic mass squared $m_s^2$ (dashed lines) in our mechanism (blue), and by contrast in hybrid inflation with a mild waterfall phase (orange), see the paragraph ``Comparison with hybrid inflation'' for details. We marked as grey the region where the dynamics is not relevant for enhancement.}} 
\label{fig:characteristics}
\end{figure}

We use this effective approach to present our PBH generation mechanism independently of any precise microscopic realization, and to focus on its main characteristics:\footnote{Concrete UV realizations of our mechanism may exhibit further features, making its phenomenology even richer.} we assume that $H(N)$ is featureless, and consider a strong turn in field space, with $\etaperp^2 \gg 1$ around some time $\Nf$, well after the CMB scales exit the Hubble radius, while being negligible before and after the turn (see fig.~\ref{fig:characteristics}). As for the entropic mass, we write $m_s^2/H^2=b- \etaperp^2$, where we consider $b$ constant for simplicity, concentrating on the time variation of $m_s^2$ set by the bending itself. An important parameter is the duration of the turn in \textit{e}-folds $\delta$, which leads to qualitatively different behaviours for broad and sharp turns ($\delta \gtrsim$ or $\lesssim \ln(\etaperp)$ respectively), to be discussed in turn below. However, these two regimes share important qualitative characteristics: 1. Scales that cross the Hubble radius well after the turn are not affected by it and their dynamics is vanilla single-field, with power spectrum $\mathcal{P}_{0}=H^2/(8 \pi^2 \epsilon \Mp^2)_{k=aH}$. 2. Scales that cross the Hubble radius well before the turn are subject to a standard multi-field mechanism: the turn results in a transfer from entropic to curvature fluctuations on super-Hubble scales (see e.g.~\cite{Bassett:2005xm}). The corresponding boost of ${\cal P}_\zeta$ compared to $\mathcal{P}_{0}$ depends (mainly) on the value of $m_s$ from Hubble crossing until the time of the turn. For definiteness, one will consider a non-negligible value of $b$, expected on general grounds \cite{Chen:2009we,0911.3380}, for which entropic fluctuations have sufficiently decayed by the time of the turn, to result in no amplification of ${\cal P}_\zeta$ on these large scales. 3. For scales that cross the Hubble radius soon after the time of the turn (a statement made more precise below), the large tachyonic mass of the entropic fluctuations results in their exponential growth, a transient instability which is concurrently transferred to the curvature perturbation by the kinetic coupling provided by the bending. The resulting exponential enhancement of ${\cal P}_\zeta$ compared to $\mathcal{P}_{0}$ on these scales is the salient observational feature of strong turns, which we now characterize in detail.

{\bf Broad turns.}--- For broad enough turns, one can appropriately adapt the results corresponding to constant large turns, which have recently been studied \cite{Garcia-Saenz:2018ifx, Garcia-Saenz:2018vqf, 1902.03221,1908.11316}. The dynamics of each $k$-mode is characterized by two times, $\tilde{N}$ corresponding to entropic mass crossing and the onset of the instability, and $\Nbar$ corresponding to the effective sound horizon crossing where $\zeta$ becomes constant:
\begin{equation}
\frac{k}{a(\tilde{N})}=|m_{s}(\tilde{N})| \, ,\qquad\&\qquad\frac{k |c_s|}{a(\Nbar)}= H(\Nbar) \, ,\label{entrcrossing}
\end{equation}
Here, $c_s^2=m_s^2/(m_s^2+4 H^2 \etaperp^2)<0$ denotes the (square of the) imaginary speed of sound, describing the transient exponential growth of fluctuations in the language of the single-field effective field theory of inflation. If background quantities vary only mildly between  between $\tilde{N}$
and $\Nbar$, which parametrically necessitates $\delta \gtrsim \ln(\etaperp)$, one can write
\begin{equation}
\mathcal{P}_\zeta(k)=\mathcal{P}_{0}(k)e^{2\,x}|_{\tilde{N}_{k}},\label{approx}
\end{equation}
for each $k$ mode that satisfies $k/a(\tilde{N}_{k})=|m_{s}|$ when $m_{s}^{2}<0$, where $x=\frac{\pi}{2}\left(2-\sqrt{3+b/\etaperp^2}\right)\eta_{\perp}$ and where the evaluation at $\tilde{N}_{k}$ is motivated by the fact that most of the growth occurs at early times.

\begin{figure}[t]
	\includegraphics*[width=9cm]{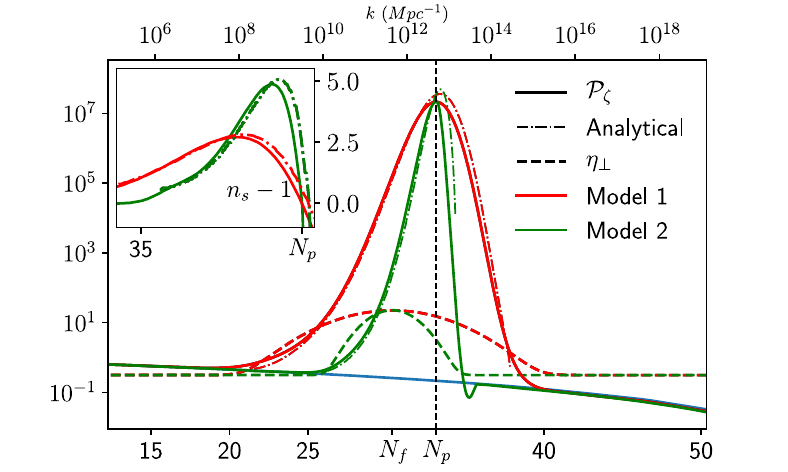}
	\caption{\textit{
		Power spectra for the two models discussed in the text compared to their bending profiles. The dot-dashed curves represent the analytical approximation in Eq.~\eqref{approx}. The blue line corresponds to the single-field power spectrum $\mathcal{P}_0$ for the same $H(N)$. Here and in other figures, power spectra are normalized to one at the CMB pivot scale.
		}}
	\label{fig:powers}
\end{figure}

Fig.~\ref{fig:powers} shows the good agreement between this analytical treatment and full numerical results, in the case of a Gaussian profile for the bending $\eta_{\perp}=\etamax e^{-(N-\Nf)^{2}/(2\sig^{2})}$. Two benchmark models have been chosen, characterized respectively by $(\etamax,\sig^2)=(22.53,10)$ and $(22.9,2)$, with $b=1$, $\Nf=30.3$ after Hubble exit of the CMB pivot scale, and $H(N)$ corresponding to Starobinsky inflation. Values for numerical parameters were chosen for later convenience, although perturbativity constraints to be discussed below are not taken into account here.

From the analytical approximation \eqref{approx}, one deduces that the mode $\kp$ corresponding to the peak of the power spectrum is the one whose entropic mass crossing coincides with the peak of $\etaperp$, i.e.~$\tilde{N}_{\kp}=\Nf$, and hence 
\begin{equation}
\kp \simeq\kf \left((\etamax)^2-b \right)^{1/2} \, ,
\label{peak}
\end{equation}
where $\kf$ is the scale that exits the Hubble radius at $\Nf$, while the corresponding height of the peak is given by
\begin{equation}
\gamma\equiv\ln\left(\frac{\mathcal{P}}{\mathcal{P}_{0}}\right)\Big|_{\textrm{peak}}=\pi(2-\sqrt{3+b/(\etamax)^2}) \, \etamax \, .
\end{equation}
In the relevant regime $(\etamax)^2 \gg b$, one can thus learn when the turned happened from the location and the amplitude of the peak.
To estimate the growth rate of the power spectrum, one can compute
the spectral index:
\begin{align}
(n_{s}-1)-(n_{s}-1)_{0}&\simeq \pi(2-\sqrt{3})\frac{\Nf-\tilde{N}}{\sig^{2}+\Nf-\tilde{N}}\,\eta_{\perp} \, ,
\label{ns}
\end{align}
where all quantities on the left hand side are evaluated at time $\tilde{N}_k$. 
This shows that the stronger and/or the less broad the turn, the steeper the peak of the power spectrum (within the regime of validity of the analytical approximation). While not surprising by itself, it is interesting that it can easily overcome the bound for $n_s$ found in single-field inflation \cite{Byrnes:2018txb, Carrilho:2019oqg,Ozsoy:2019lyy}. The very good agreement between the analytical formula for $n_s$ and the numerical result is shown in fig.~\ref{fig:powers}, where one can see that $n_s-1$ reaches $5$ in the model with a sharper bending.

\begin{figure}
  \includegraphics*[width=9cm]{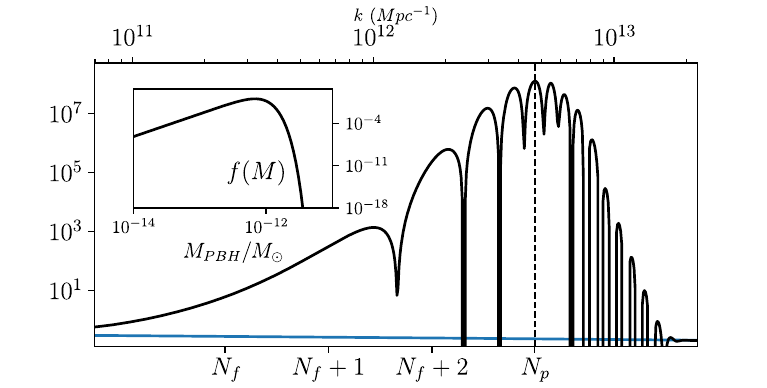}
  \caption{\textit{For sharper bending ($\eta_{\perp}^{\mathrm{max}}=68$, $\sig^2=0.1$) the shape of the power spectrum exhibits characteristic oscillatory patterns. These oscillations do not translate into features of the PBH mass distribution, at least with the assumption of a Gaussian probability distribution function.}}
\label{fig:oscillations}
\end{figure}

{\bf Sharp turns.}--- When the turn is shorter, with a duration $\delta \lesssim \log(\etaperp)$, the scales that are maximally enhanced, with entropic mass crossing during the turn, and hence still of order $k \sim \kf \etamax$, are well inside the Hubble radius when the turn ends, with the non-trivial dynamics during the turn effectively generating an excited initial state for these modes. This can be studied analytically in the regime of sharp turns with $\delta \ll 1$ (and constant $\etaperp$ during the turn) \cite{Fumagalli:2020nvq,Palma:2020ejf}\footnote{Ref.~\cite{Palma:2020ejf} appeared at the same time as the first version of this work.}: in addition to a localised exponential enhancement of the (envelope of the) power spectrum, the latter is modulated by rapid order one oscillations in $k$, with frequency $2/ \kf$ set by the time of the turn, and which are characteristic of sharp features \cite{Chluba:2015bqa,Slosar:2019gvt}. More realistic time-dependences of $\etaperp$ during the turn, like our Gaussian profile, also display these patterns, as can be seen in fig.~\ref{fig:oscillations}.

{\bf Comparison with hybrid inflation.}--- Hybrid inflation with a mild waterfall phase is a multi-field scenario that is well known to be able to generate a substantial amount of PBHs \cite{1501.07565}, and it is instructive to compare and contrast the mechanism at play there with the one presented here. For this, we show in fig.~\ref{fig:characteristics}, for the two setups, the characteristic time-dependences of the coupling $\etaperp$ and the entropic mass $m_s^2/H^2$ entering the action \eqref{L2} describing the linear fluctuations. Hybrid inflation also displays a turn with a temporary (modest) increase of $\etaperp$, but this is preceded by a period lasting many \textit{e}-folds (the mild waterfall phase) during which $m_s^2/H^2 \sim -1$ is mildly tachyonic. During that stage, super-Hubble entropic fluctuations grow, effectively decoupled from the curvature perturbation, before transferring their power to the latter during the turn, which happens at the same time that the instability shuts off. 
Hence, in hybrid inflation, the boost of the curvature power spectrum stems from a two-stage process and is the result of a standard super-Hubble classical phenomenon. In contrast, in our scenario with a strong turn, the growth of entropic fluctuations and their transfer to the observable curvature power spectrum occur at the same stage -- the two phenomena cannot actually be disentangled -- as the large $\etaperp$ controls both the coupling between perturbations and the large negative $m_s^2/H^2$. As a result, the boosted power spectrum in our scenario results from a genuinely quantum phenomenon arising before Hubble crossing. These differences are particularly striking in the limit of sharp turns, leading to oscillatory features that can not be mimicked by classical phenomena and that induce characteristic observational signatures in GWs \cite{Fumagalli:2020nvq,Braglia:2020taf,Fumagalli:2021mpc}.


{\bf PBH spectrum.}---
The main quantity of interest is the mass spectrum $f(M)$, which is related to the fraction of DM in PBHs through $\Omega_{\textrm{PBH}} = \Omega_{\textrm{CDM}} \int f(M) \textrm{d} \ln M$. The relevant quantity that determines whether a region collapses into a PBH is the smoothed density contrast \cite{1405.7023,1905.01230} and for computations we use the definitions in eqs.~(14)-(17) of \cite{1405.7023}. Here we assume that fluctuations in the density contrast induced by the primordial curvature fluctuations are Gaussian and that PBH formation occurs during a period of radiation domination. We will remark on the potential effects of NG later. The computation of $f(M)$ is then standard and we use eq.~(26) in \cite{1801.06138} (see \cite{Suyama:2019npc} for recent improvements).\footnote{This formula for $f(M)$ depends on several numerical parameters whose values are taken from simulations and we use $\mu \equiv M / (C \, M_H)$ with $C=3.3$, $\gamma=0.36$, $\delta_c = 0.45$ following \cite{9709072, 9901292, 0412063, 0811.1452, 1201.2379}. To relate the comoving scale $k$ of a mode re-entering the horizon to the corresponding Hubble volume mass $M_H$ we evolve these quantities from their respective values at matter-radiation equality, using $k = k_{\textrm{eq}} \sqrt{M_{H,\textrm{eq}} / M_H}$ with $k_{\textrm{eq}} =0.01 \, (\Omega_m / 0.31) \, \textrm{Mpc}^{-1}$ and $M_{H, \textrm{eq}} \approx 2.8 \times 10^{17} M_{\odot}$.} 

\begin{figure}
  \includegraphics*[width=9cm]{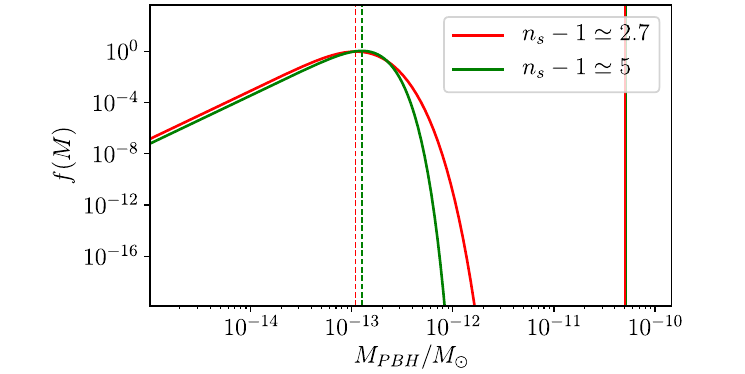}
\caption{\textit{PBH mass spectrum $f(M)$ for the two benchmark models, corresponding
to $n_s-1= 2.7, 5$. The separation in mass between the peak of $f(M)$, indicated by the dashed lines, and the horizon mass corresponding to the scale of maximal bending, denoted by the solid lines, is a direct consequence of Eq.~\ref{peak}.}}
\label{fig:massfraction}
\end{figure}

In fig.~\ref{fig:massfraction} we plot $f(M)$ for the two benchmark models for a broad turn and in the inset of fig.~\ref{fig:oscillations} for the sharp turn example. The model parameters have been chosen purposefully such that PBHs constitute all of DM at matter-radiation equality, i.e.~$\Omega_{\textrm{PBH}} = \Omega_{\textrm{CDM}}$, and so that $f(M)$ peaks near $M_{\textrm{PBH}} \sim 10^{-13} M_{\odot}$, which lies in the window where PBHs can constitute a significant fraction of DM \cite{2002.12778}. For the two broad turn examples we observe that a steeper growth of the power spectrum corresponding to a larger value of $n_s$ results in a narrower peak for $f(M)$. Interestingly, the oscillations in $\mathcal{P}_\zeta$ observed for the sharp turn case do not lead to obvious visible features in $f(M)$: When computing the smoothed density contrast, which for the maximally enhanced scales amounts to averaging over a scale comparable to the width of the peak in $\mathcal{P}_{\zeta}$, the oscillations are smoothed out.



{\bf Discussion.}--- Modern embeddings in high-energy physics suggest that inflation may be of multi-field type. In this context we observe that periods during which the inflationary trajectory strongly deviates from a geodesic in field space can strongly enhance scalar fluctuations, in a functionally different way from previous multi-field mechanisms, and with the effect on the scalar power spectrum $\mathcal{P}_{\zeta}$ markedly different for broad vs.~sharp turns.

A challenge faced by every mechanism that enhances quantum fluctuations is that these may backreact excessively on the background, or that their perturbative treatment may become invalid. Tentative bounds can be given based on the assessment that perturbative control is maintained if higher-order interactions are suppressed compared to the quadratic Lagrangian, $\mathcal{L}_n / \mathcal{L}_2 < 1$. This gives $\etaperp^6 \mathcal{P}_\zeta < 1$ for broad \cite{1902.03221} and $\etaperp^4 \mathcal{P}_\zeta < 1$ for sharp turns \cite{Fumagalli:2020nvq}. The bounds for avoiding excessive backreaction are similar, and typically weaker \cite{Fumagalli:2020nvq}. According to these criteria our benchmark parameter choices (which were mainly chosen for illustrative reasons) are in danger, but this is not conclusive from these tentative bounds. Instead, this highlights the importance of these constraints, which should be analysed for every enhancement mechanism (see also \cite{Atal:2018neu, Adshead:2014sga, Inomata:2021uqj}), and rather motivates further research on more precise formulations of these bounds.

Another important question for further investigation concerns the effect of NG on the probability distribution of density fluctuations and hence the PBH abundance \cite{1206.4188,1307.4995,1801.09415,1811.07857,DeLuca:2019qsy,1906.02827,1906.06790,1908.11357,1912.05399}. Inflationary trajectories with strongly non-geodesic motion exhibit a characteristic pattern of NG, with a bispectrum and higher-order correlation functions enhanced for flattened configurations \cite{Garcia-Saenz:2018vqf, 1902.03221, 2003.13410}, whose effect may be to either suppress or boost the PBH abundance compared to the Gaussian ansatz. The latter situation is particularly interesting as it may allow for a significant PBH production for lower values of ${\cal P}_\zeta$, hence alleviating the theoretical constraints above. Another exciting prospect is that this type of NG may manifest itself in a distinct feature in $f(M)$. An additional pathway for experimental scrutiny of our mechanism is via its GW signal, which has been analysed in a more general context in \cite{Fumagalli:2020nvq,Braglia:2020taf,Witkowski:2021raz,Fumagalli:2021mpc} for sharp turns. The oscillations in $\mathcal{P}_\zeta(k)$ are reprocessed into corresponding oscillations in the GW energy fraction spectrum $\Omega_{\textrm{GW}}(f)$ accentuating the rich phenomenology of this mechanism.


{\bf Note added.}--- While the first version of this work was being finalized, a similar idea appeared in \cite{Palma:2020ejf}. Where overlapping, our results agree.


\begin{acknowledgments}

We are grateful to Sebastian Garcia-Saenz, Shi Pi, Lucas Pinol, Caner \"Unal and Vincent Vennin for interesting and helpful discussions, as well as the anonymous referees who helped improving the paper. J.F, S.RP, J.W.R and L.T.W are supported by the European Research Council under the European Union's Horizon 2020 research and innovation programme (grant agreement No 758792, project GEODESI).

\end{acknowledgments}


\begin{center}
\textbf{\large Appendices}
\end{center}
\setcounter{table}{0}
\makeatletter

\subsection{From broad to sharp turns: the appearance of the oscillations}
\setcounter{equation}{9}
\setcounter{figure}{4}
Oscillations in the primordial power spectrum occur for enhanced modes that are still well inside the Hubble radius when the turn ends. As remarked in the main text, this is a critical difference compared to previous proposals such as the mild waterfall phase of hybrid inflation. In this supplemental material, we spell out the conditions for these oscillations to appear. For simplicity, let us first focus on the simple case of a top-hat profile for the time dependence of the bending parameter so that $\etaperp(N)$ has width $\delta$, height $\etaperp$ and is centered at $\Nf$. The enhanced modes are approximately those whose entropic mass crossing happens between $N_1 \equiv \Nf - \delta/2$ and $N_2\equiv \Nf +\delta/2$, i.e.~the times when the turn starts and ends respectively.
In particular, we denote a turn as sharp if all enhanced modes are still sub-Hubble at $N_2$, i.e.~$N_1 +\ln\etaperp \gtrsim N_2$. This gives back the criterion $\delta \lesssim \ln \etaperp$ used in the main text. The inequality being saturated means that the first enhanced mode, i.e.~the one for which $\tilde{N} = N_1$, becomes super-Hubble just at the end of the turn $N_2$, where we remind that $\tilde{N}$ denotes evaluation at entropic mass crossing such that $k=a |m_s|$. In that situation, all other enhanced modes are (mildly) sub-Hubble at $N_2$ and so not yet frozen at the end of the turn ---see figure \ref{differentmodes}.  
From that time onwards, the enhanced modes are effectively in an excited initial state \cite{Fumagalli:2020nvq} and, as a consequence, the primordial power spectrum displays characteristic oscillations at these scales ---see figure \ref{transition}.  
\begin{figure}
	\includegraphics*[width=7cm]{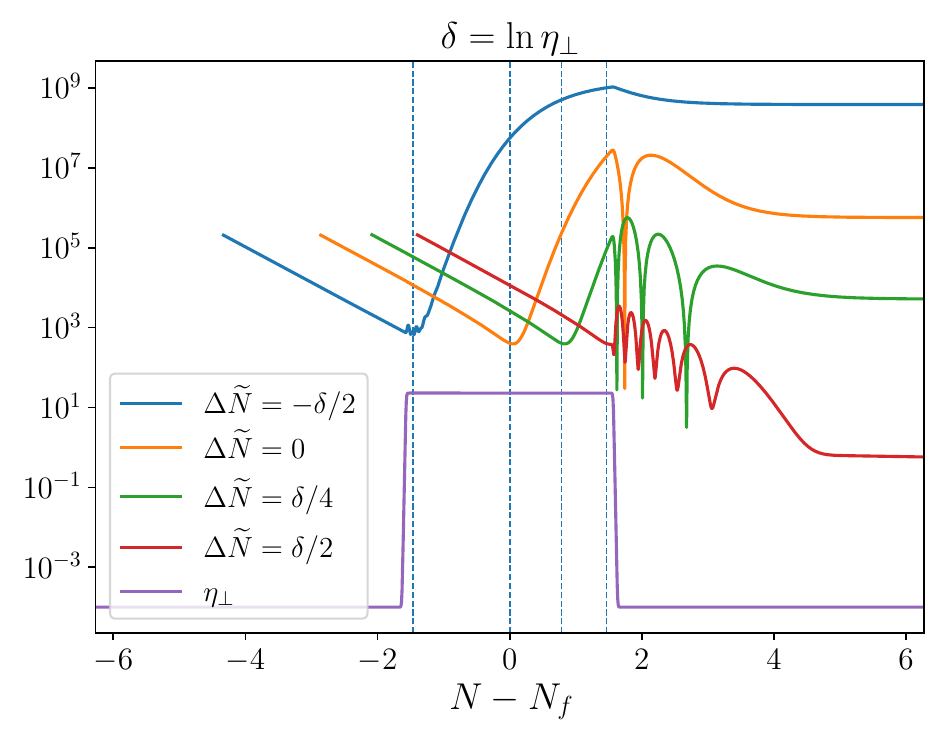}
	\caption{ \textit{Time dependences of the power spectra for $k$ modes with different entropic mass crossing $\Delta \tilde{N}\equiv \tilde{N} - N_f$, for the top-hat time-dependence of $\eta_\perp$ discussed in the text. 
$\delta = \ln \etaperp$ is the limiting situation between sharp and broad turns. In this case, the first (smallest k) enhanced mode crosses the Hubble radius at the end of the turn. Thus, all other enhanced modes are sub-Hubble at the end of the turn.}}

\label{differentmodes}
\end{figure}
\begin{figure}
	\includegraphics*[width=7cm]{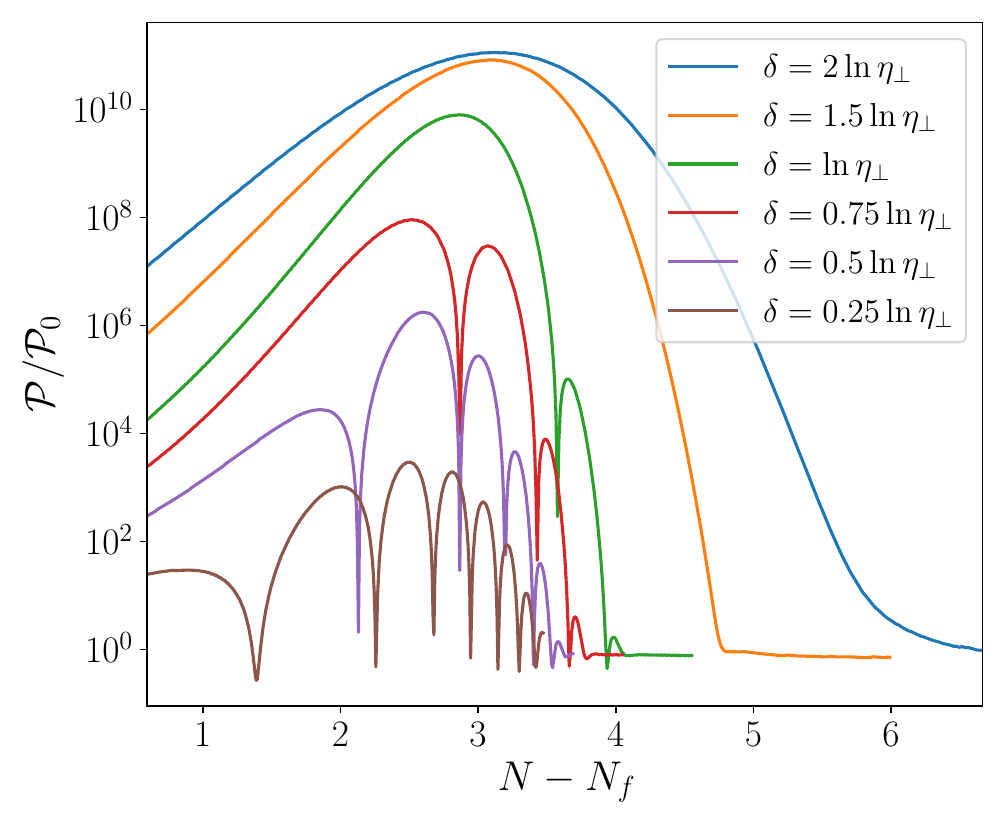}
	\caption{ 
		\textit{Examples of power spectra for increasingly squeezed Gaussian profiles for the turn, labeled by the parameters $(\etaperp,\delta)$, which are related to  $(\etamax,\Delta)$ through the dictionary explained in the text. We fix $\etaperp = 25$ and consider a set of decreasing values of $\delta$.  
		Oscillations start to appear at the transition $\delta \simeq \ln \etaperp$ because enhanced modes are not yet super-Hubble at the end of the turn.}}
\label{transition}
\end{figure}

Let us now consider the more realistic case of a smooth profile such as the one of a Gaussian used in the main text, i.e.~$\eta_{\perp}(N)=\etamax e^{-(N-\Nf)^{2}/(2\sig^{2})}$. We identify, for instance, the duration of the turn with $\delta = 3\sig $ and as the typical value for the bending parameter we define a constant $\etaperp$ such that $\etaperp \delta = \int \etaperp (N)dN = \sqrt{2\pi \sig^2} \etamax$. This is motivated by the fact that the enhancement of the power spectrum is sensitive to the total angle swept in field space.\footnote{Upon the identification $(\etaperp,\delta) \leftrightarrow (\etamax, \sig)$ the two benchmark examples for broad turns in the main text have $(\ln \etaperp, \delta) \simeq (3, 9.5)$ (Model 1) and $(\ln \etaperp, \delta) \simeq (3, 4.24)$ (Model 2). So in both cases $\delta \gtrsim \ln \etaperp$.} Upon making these identifications, figure \ref{transition} shows the transition between the two regimes $\delta \gtrsim \ln \etaperp$ and $\delta \lesssim \ln \etaperp$ by keeping $\etaperp$ fixed and varying $\delta$. For large enough $\delta$, the last enhanced mode is already super-Hubble when the effect of the bending ends, which results in a smooth power spectrum. As $\delta$ is reduced, visible oscillations in $\mathcal{P}_\zeta$ appear. Those arise first on the right-hand side of the peak and gradually proceed to the left, since reducing the duration of the turn implies that lower $k$ modes become sub-Hubble at the end of the turn. Remarkably, and contrary to the standard lore, the presence of oscillations
does not necessarily require the turn to be very sharp, i.e.~$\delta \ll 1$. 
For instance, in the example of figure \ref{transition} ($\etaperp = 25$), oscillations are clearly visible for a turn lasting approximately $\delta = 0.75 \cdot  \ln 25 \simeq 2.4$ $e$-folds.

\subsection{Shape of the power spectrum}
\begin{figure}[t]
	\includegraphics*[width=9cm]{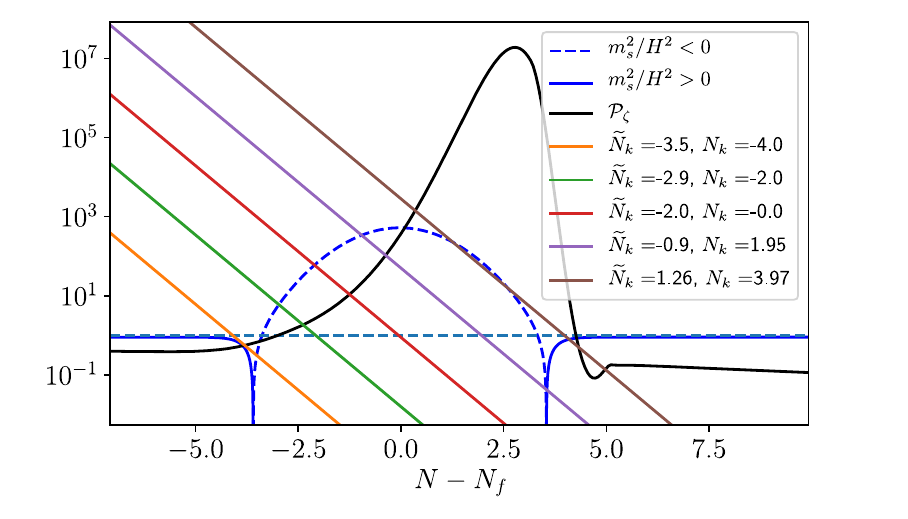}
	\caption{ 
		\textit{Family of lines $k^2/(aH)^2$ equally $\ln k$-divided, with $ \delta N \simeq 2$, for modes whose entropic mass crossing happens while $m_s^2<0$
		for Model 2. More modes enter the region of instability before the parameter $|m_s^2|/ H^2$ has passed its peak than after, resulting in an asymmetric peak for the power spectrum with a sharp falloff.}}
\label{texplanation}
\end{figure}
\noindent 
The mechanism proposed in this Letter leads to a characteristic shape of the scalar power spectrum that can potentially leave observational imprints in the PBH mass distribution and/or in the profile of the scalar-induced stochastic gravitational wave background. In this supplemental material we highlight that a strong turn in field space naturally results in an enhanced power spectrum that falls abruptly after its peak, a statement that holds for any duration of the turn. Moreover, for broad turns, we provide an expression for the growth of the power spectrum that is independent of the profile of the turn.

To see explicitly the sharp falloff of the power spectrum after its peak, we begin by approximately estimating the last mode that is enhanced.
Let us remind that the modes enhanced are the ones whose entropic mass crossing happens while $m_s^2<0$. From fig.~\ref{texplanation}, one can hence intuitively understand the asymmetry of the peak, as a larger range of modes enjoys this characteristic before the peak of $|m_s|$ (entropic mass crossing at this point corresponding to the peak of the power spectrum) than afterwards.  
More quantitatively, one can identify the largest mode $\ke$ to be enhanced as the one whose line $k^2/(aH)^2$ in fig.~\ref{texplanation} is tangent to $m_s^2/H^2$, i.e. the one verifying 
\begin{equation}
\ln|m_{s}(N)|'\Big|_{N=\tilde{N}_{\mathrm{e}}}=-1.
\end{equation}
For the case of a Gaussian profile for the bending,
this gives
\begin{equation}\label{ss}
\frac{(\tilde{N}_{\mathrm{e}}-\Nf)/\sig^{2}}{1-b/\eta_{\perp}^{2}(\tilde{N}_{\mathrm{e}})}=1, 
\end{equation}
and the corresponding \textit{e}-fold at Hubble crossing is given by 
\begin{equation}\label{Ne}
N_{\mathrm{e}}=\tilde{N}_{\mathrm{e}}+\ln\left(\tfrac{|m_{s}(\tilde{N}_{\mathrm{e}})|}{H}\right).
\end{equation}
Let us see how this argument applies, for the Gaussian profiles of our two benchmark models for a broad turn: $(\etamax,\sig^2)=(22.53,10)$ and $(22.9,2)$; and of our model 3 of figure \ref{fig:oscillations} for a sharp turn: $(\etamax,\sig^2)=(68,0.1)$ (all with $b=1$). By numerically solving \eqref{ss} and plugging the results back into \eqref{Ne} we obtain $N_{\mathrm{e}}-N_{\mathrm{p}} \simeq 4.57 \,\,(\simeq 0.99 / \simeq 1.15 )$ for model 1 (model 2/ model 3). The three values are in agreement with the falloff of the power spectrum observed in figure \ref{texplanation}, \ref{fig:powers} and \ref{fig:oscillations}, where the asymmetry of the peak of the power spectra is also manifest. Moreover, comparing models with approximately the same value of $\etamax$, here models 1 and 2, one finds that the sharper the bending profile the steeper the falloff of the power spectrum. 
As an aside, note that as one considers turns with $(\eta_\perp^{\textrm{max}})^2 \gg b$, one may be tempted to solve eq~\eqref{ss} by considering the denominator there to be almost unity, trivially finding $\tilde{N}_{\mathrm{e}}\simeq \Nf+\sig^{2}$, and hence $N_{\mathrm{e}}-N_{\mathrm{p}} \simeq \sig^2/2$ for broad turns. However, this simple solution is self-consistent only under the restrictive condition $b/(\etaperp^{\mathrm{max}})^{2}\exp(\sig^{2})\ll 1$. This is verified in model 2, with this parameter being $\simeq 0.01$, but not for model 1 with a broader turn.

As discussed in the main text (and in the previous supplemental material), the primordial power spectrum does not display oscillations if the modes that are enhanced cross the Hubble radius before the end of the turn, in which case we label the turn as broad.
In particular, all enhanced modes are already super-Hubble by that time if $N_{\mathrm{e}} \lesssim N_2 $, where $N_2$ is the time where the turn is approximately over.\footnote{Note that as an artefact of choosing a top hat profile for $\etaperp$ we have that $N_e \lesssim N_2$ is never satisfied in that case. That means that no matter how broad the top hat is, there will always be a range of modes $\tilde{N}\in [N_2-\ln \etaperp, N_2]$ that will exhibit oscillations in $\mathcal{P}_\zeta$.}
Numerically solving for $N_e$ as a function of the parameters of the turn, one can check that this is indeed parametrically equivalent to the condition $\delta \gtrsim \ln \etaperp$ previously stated.

Finally, let us derive a general expression for the growth of the primordial power spectrum that is valid for broad turns, independently of the precise time-dependence of its profile.
By carefully differentiating Eq.~\eqref{approx} one obtains
\begin{align}
(n_{s}-1)-(n_{s}-1)_{0}=K\frac{d \eta_{\perp}}{d\tilde{N}}\left(1+\frac{d\ln\eta_{\perp}}{d\tilde{N}}\frac{1}{(1-b/\eta_{\perp}^{2})}\right)^{-1},
\label{ns2}
\end{align}
with $K = \pi(2-\sqrt{3})$, all background quantities on the right-hand side are evaluated at the time $\tilde{N}$, and which reduces to \eqref{ns} for a Gaussian profile.
As explained in the main text, either the generic formula above or Eq.~\eqref{ns} tell us that, within their regime of applicability, sufficiently strong or/and less broad turns can easily make $n_s$ large so to overcome the single-field inflation bound for $n_s$ found in \cite{Byrnes:2018txb, Carrilho:2019oqg,Ozsoy:2019lyy}.

Note that the expression for the spectral index just discussed is derived from the analytical approximation in Eq.~\eqref{approx} for the primordial power spectrum. Let us discuss then in more detail its regime of validity: Eq.~\eqref{approx} can be used if the quantity $x(N)$ can be considered approximately constant between $\tilde{N}$
and $\Nbar$. Namely, for a Gaussian profile  ~$(\Nbar-\Nf)^{2}-(\tilde{N}-\Nf)^{2}\lesssim 4(\sig^{2}$).\footnote{For the mode with $\tilde{N}=\Nf$, the criterion for the validity of the approximation reads $(4 \sig^2)^{-1}\ln^2(|c_s|\etamax) \lesssim 1$, with the left-hand side given by $0.1$ and $0.7$ for Model 1 and 2 respectively.}
In general, one expects this approximation to underestimate the power spectrum before the peak, as $\eta_{\perp}$
grows in the interval $[\tilde{N},\Nbar]$ for those modes, and to overestimate it after the peak for opposite reasons. This is indeed what we observe in fig.~\ref{fig:powers}, but as previously shown, this does not 
prevent our analytical scheme to provide relevant information. Note as well that we have checked that Eq. \eqref{ns2} provides a good estimate of the growth of the power spectrum even in the intermediate regime $\delta\simeq O(1) \ln \etaperp$ where oscillations are already present in the falloff part of the peak.

 \bibliographystyle{apsrev4-1}
\bibliography{PBHbib}

\end{document}